\documentstyle[12pt,epsf]{article}

\begin{document}

\title{Griffiths singularities in the two dimensional diluted Ising model}
\author{Juan J. Ruiz-Lorenzo\\[0.5em]
{\small  Dipartimento di Fisica, Universit\`a di Roma}
   {\small {\em La Sapienza} }\\
{\small and}\\
{\small Istituto Nazionale di Fisica Nucleare, Sezione di Roma I}\\
{\small   \ \  Piazzale Aldo Moro 2, 00185 Roma (Italy)}\\[0.3em]
{\small \tt ruiz@chimera.roma1.infn.it}\\[0.5em]}

\date{\today}

\maketitle

\begin{abstract}

We study numerically the probability distribution of the Yang-Lee
zeroes inside the Griffiths phase for the two dimensional site diluted
Ising model and we check that the shape of this distribution is that
predicted in previous analytical works. By studying the finite size
scaling of the averaged smallest zero at the phase transition we
extract, for two values of the dilution, the anomalous dimension,
$\eta$, which agrees very well with the previous estimated values.

\end{abstract}  

\thispagestyle{empty}
\newpage

%123456789%123456789%123456789%123456789%123456789%123456789%123456789%1
\section{\protect\label{S_INT}Introduction}

The Yang-Lee theorem provides a theoretical, and powerful, tool to
study phase transitions. In systems without disorder ({\it e.g.} the
usual $\phi^4$ theories or Ising models, XY model, etc)
this theorem allows to characterize and to estimate numerically the
phase transition and the anomalous dimension \cite{GENERIC,KENNA}.

In the disordered case ({\it i.e.} systems with a random interactions)
the theorem provides a tool to study (and to define) the Griffiths
phase (or in other words the Griffiths singularities)
\cite{GRIFFITHS}.  The Griffiths phase is a peculiar phenomenon of
disordered systems. Roughly, it is a region above the critical
temperature of the disordered system and below that of the pure system
(for some choices of the disorder distribution this temperature could
be infinite \cite{BRAY}).  Below the critical temperature of the pure
system, which we denote $T_c(p=1)$, but above the critical temperature
of the disordered one, which we denote $T_c(p)$, there exist
magnetized domains (geometrical clusters, since of course, the total
magnetization is zero, as we are still in the paramagnetic phase of the
diluted system). These domains of non-zero magnetization induce a
complex singularity (Yang-Lee zeroes) in the free energy as a function
of the magnetic field (Griffiths singularity \cite{GRIFFITHS}).

In  classical statistical mechanics the Griffiths singularities are 
essential singularities and so have no effect on the static
properties of the system (nothing diverges in the Griffiths phase,
except at the critical point\footnote{
    In the quantum case the singularities are stronger \cite{RIEGERYOUNG}.
}). 

But dynamically this phase induces a
slow behavior in the spin-spin autocorrelation functions \cite{OGIELSKY}
, the dynamic
of the system becomes slower than in the ``usual'' paramagnetic phase
\cite{BRAY}.

For instance, in the three dimensional spin glass case, numerically
there is a change in the autocorrelations functions from those of
the paramagnetic case ($C(t) \sim t^{-x}\exp(-a t)$) to a short range
correlations (like a behavior \footnote{It is possible to demonstrate
rigorously that for Ising like models (diluted, spin glasses, etc) the
behavior must be: $C(t) \sim \exp(-a (\log t)^{d/(d-1)})$. I thank
F. Cesi for pointing  this fact to me \cite{FILIPPO}.}  : $C(t) \sim
t^{-x}\exp(-a t^\beta),\;\;\beta \neq 1$ ) just at the critical point
of the {\em pure} Ising model. Obviously at the critical point of the
$3d$ spin glass there exists another change in the behavior of the
autocorrelation function to a spin glass regime \cite{OGIELSKY}.

In this paper we will focus our attention on the probability
distribution of the smallest zero in the Griffiths phase and we will
confront our numerical results with the analytical prediction of
reference \cite{BRAY}. We have obtained a clear numerical picture about
the construction of the Griffiths singularities.

We will also extract, using the scaling of the average of the smallest
zeroes at the critical point, the anomalous dimension of the system
and we will compare this value with previous numerical simulations of
the system \cite{KIMPA}.

%123456789%123456789%123456789%123456789%123456789%123456789%123456789%1
\section{\protect\label{S_YLB}Yang-Lee singularities}

By regarding the partition function of the pure Ising model in a finite
volume $L^d$ as a function of the variables 
$$
\rho=e^{-2 h} \;\;,\;\;\; \tau=e^{-2 \beta},
$$
where $h$ is the magnetic field and $\beta$ is the inverse of the
temperature, Yang and Lee \cite{YANGLEE,ID} 
found that the complex zeroes of the
partition function in the $\rho$ variable lie in the unit circle and
there are no zeroes on the real axis. Moreover in the thermodynamical
limit, and for $\beta \ge \beta_c$, the point $\rho=1$ becomes an
accumulation point giving rise to a singularity in the free energy.

Near the critical point, in the paramagnetic phase, the imaginary part of 
the zero nearest to the real axis, $h_s$,
behaves
\begin{equation}
h_s \sim  (\beta_c-\beta) ^{\Delta} ,
\end{equation}
and then, in the standard way, we can write down the finite--size
dependence of $h_s$ at the critical point
\begin{equation}
h_s \sim L^{-(\Delta/\nu)} .
\end{equation}
Using  the scaling relation $\Delta=\nu d -\beta$, where $d$ is the 
dimension, we can rewrite 
the last equation as
\begin{equation}
h_s \sim L^{-(2-\eta/2)} .
\label{CRITICAL}
\end{equation}
Below the phase transition, in the ferromagnetic phase, the scaling law is
\begin{equation}
h_s \sim L^{-d} .
\label{FERRO}
\end{equation}

In the disordered case, each sample will have a smallest $h$, that we
hereafter denote as $h_\epsilon$. We will investigate numerically the
functional form of the probability distribution of $h_\epsilon$, that
we will write as $p(h_\epsilon)$.

There are some analytical results about the density of the zeroes in the
Griffiths phase. The authors of reference \cite{BRAY}  obtain for the 
density of zeroes, with imaginary part $h_i$,
 of a diluted Ising system with a proportion of spins $p$ 
the following  law 
\begin{equation}
\rho(h_i) \propto \exp(\frac{A \log(p)}{ h_i}) ,
\label{bray_law}
\end{equation}
as $h_i \ll 1$, which is a very weak dependence. 

It has been
assumed that a cluster of size $L$ introduces a
zero, which induces the previous law, that scales as (see
equation (\ref{FERRO}))
\begin{equation}
h_i=\frac{A}{ L^{d}}.
\label{FERROMAG}
\end{equation}
where $A$ is the inverse of the site magnetization of the cluster \cite{BRAY}.

It is possible to obtain a better estimate of the prefactor of $1/h_i$
in the exponential of  the formula (\ref{bray_law}) using a
variational method \cite{BRAY}.

The important point is that there is a
finite probability to have a zero in any neighborhood of $h=0$.

To complete this discussion we will add that at the critical point
the density arrives with a non zero slope to the origin, in the
broken phase the density at the origin is finite, and above of the
critical temperature of the pure system the density is zero in a
neighborhood of the origin \cite{BRAY}.

%123456789%123456789%123456789%123456789%123456789%123456789%123456789%123
\section{\protect\label{S_MODEL}The model and the numerical method}

The simplest disordered system is the diluted Ising model. This model
describes, for instance, the Anderson localization \cite{GIORGIO}, and
has been studied analytically (using the mapping to a O($N$) theory
with cubic anisotropy in the limit $N \rightarrow 0$)
\cite{GIORGIO,CARDYKANE} and numerically \cite{PARU,Heuer,KIM,KIMPA}.

The Hamiltonian of the two dimensional site diluted Ising model in a 
hypercubic lattice of size $L$ with periodic boundary conditions is
\begin{equation}
{\cal H}_\epsilon=-\sum_{<ij>} 
\epsilon_i \epsilon_j \sigma_i \sigma_j ,
\end{equation}
where $<ij>$ denotes nearest neighbors pairs, $\sigma_i$ are the usual
spin variables and $\epsilon_i$ are  independent quenched noises which
are  $1$ with probability $p$ and $0$ with probability $1-p$. Obviously
the system will have a phase transition only if $p \ge p_c$ where
$p_c$ is the percolation threshold for the $d$-dimensional site
percolation.  For instance, in two dimension $p_c=0.592746$
\cite{STAUFFER}. 

There are analytical results for this model mainly by Dotsenko and
Dotsenko, and Shalaev \cite{Dotsenko} (DDS) 
using Renormalization Group techniques.
There is a change in the functional form of the specific heat (from
$\log|t|$ to $\log[1+ a \log|t|]$, where $|t|$ is the reduced critical
temperature and $a$ is a constant), but there is no change in the $\nu$
exponent. This result must hold for a lower dilution of spins.
For this weak disorder there are
numerical results that support this picture \cite{Heuer}.

But the authors of reference \cite{KIMPA} claim that the specific heat
follows the prediction of (DDS) but only for a lower degree of dilution, 
moreover they found  a dependence of the $\nu$
and $\gamma$ exponents with the dilution such that the $\eta$ exponent
is constant (we remark that $\gamma/\nu=2-\eta$). 
 
The end-point of the critical line (in the plane
$(\beta,p)$), $(\beta=\infty, p_c)$\footnote{
        This is the two dimensional site percolation phase transition.}, 
has  critical exponents $\nu=4/3$ and $\gamma=43/18$
which implies $\eta=5/24 \approx 0.2083$ \cite{STAUFFER}. 

The partition function for a  purely imaginary magnetic field, $ih$, in a
$d$-dimensional lattice of size $L$ is 
\begin{equation}
{\cal Z}(\beta,h) =\sum_{[\sigma]} \exp (\beta \sum_{<ij>} \sigma_i
\sigma_j 
+i h \sum_i \sigma_i ) .
\end{equation}
By defining $M=\sum_i \sigma_i$, the total magnetization of the
system, we obtain
\begin{equation}
{\cal Z}(\beta,h)=\langle\cos(h M)\rangle + i \langle\sin(h M)\rangle ,
\end{equation}
where the average $\langle(\cdot\cdot)\rangle$ is taken with ${\cal
Z}(\beta,h=0)$, {\it i.e.} a real measure.  In the paramagnetic phase
all the odd moments of the magnetization vanish, which implies 
$\langle\sin(h M)\rangle=0$ and the only singularities of the free
energy ($\log {\cal Z}(\beta,h)$) will arise from the zeroes of
$\langle\cos(h M)\rangle$.

This is the scenario for the pure systems. In the diluted case we need
to replace $\sigma_i$ by $\epsilon_i \sigma_i$ and so each samples
will have its own smallest zero ($h_\epsilon$).  The averaged values
over all the samples, ${\overline{h_\epsilon}}$, should follow the
previous finite--size scaling relation (\ref{CRITICAL}) at
criticality.

%123456789%123456789%123456789%123456789%123456789%123456789%1234567
\section{\protect\label{S_DIS}Probability distribution of the Yang-Lee 
zeroes in the Griffiths phase}

To check the analytical form of the probability 
distribution, $p(h_\epsilon)$, of the
smallest Yang-Lee zeroes\footnote{
        Obviously in $\rho(h)$ are all the possible zeroes, but as 
        we are interested in the $h \ll 1$ regime then 
        $p(h) \approx \rho(h)$.}
we have done numerical simulations with  $\beta=0.52$ and  $p=0.889$
which is inside of the Griffiths phase.\footnote{
        We remark that for this dilution the
        phase transition is at $\beta=0.5380(3)$ 
        and the phase transition of
        the pure model is at $\beta=\frac{1}{2} 
        \log(1+\sqrt{2})=0.44069$.}
We used the Wolff algorithm \cite{WOLFF} and we simulated the 
sizes $L=4$ (8000 samples), $L=8$ (15000 samples), 
$L=12$ (2200 samples) and $L=16$ (3926
samples). The results are shown in Figure 1.
\begin{figure}[htbp]
\begin{center}
%\addvspace{1 cm}
\leavevmode
\epsfysize=250pt
\epsffile{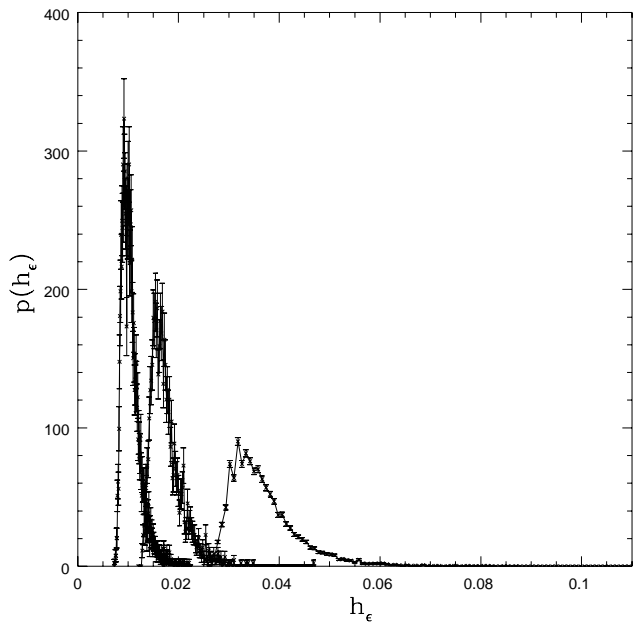}
\end{center}
\caption{Probability distribution of the smallest zeroes for (right to
left) $L=8$, $L=12$ and $L=16$. }
%\label{fig:chi}
\end{figure}

We expect that the minimum value of $h_\epsilon$, at fixed $L$, should be 
due to the sample with all the points filled ({\it i.e.} a pure Ising
model of size $L$). We find that the minimum  smallest zero, as a
function of the lattice size, follow the rule
\begin{equation}
 h_\epsilon^{\rm min} = \frac{1.5(1)}{L^{1.93(4)}},
\label{scal}
\end{equation}
where we have fitted using $4 \le L \le 12$ with 
$\chi^2/{\rm DF} =0.2/1$ (DF means  degrees of freedom).

For  $L\ge 16$ lattices the previous fit (\ref{scal}) does not hold.  
This discrepancy comes  from the fact that the number of
samples that we need to pick up this minimum value is larger than the
number that we have simulated. 

Simulating directly the pure Ising model we find  that the smallest
zero (simulating up to $L=32$), as a function of the sizes, 
at $\beta=0.52$ (ordered phase of
the pure model) behaves
\begin{equation}
h^{\rm min}_{\rm pure} = \frac{1.55(4)}{L^{1.97(1)}},
\label{thefit}
\end{equation}
following very well  the law (\ref{FERRO}).
The agreement with eq. (\ref{scal}) is also  very good.

We have also fitted the mean value of the probability distribution as
a function of $L$ (for $L$=$8$, $12$, $16$, $64$, $128$, $192$)
in the diluted case
and we have found that the numerical data behave
$$
\overline{h_\epsilon}= 0.0041(1)+ 1.67(2) L^{-1.84(1)},
$$
with a very good $\chi^2/{\rm DF} =4.9/4$. The samples used for $L \ge 64$ 
are written in Table 1.

We plot in Figure 2 the head ({\it i.e.} 
the region of lower values of $h$) of
the probability distribution for the $L=8$ lattice in the variables
$(1/h_\epsilon, \log p(h_\epsilon))$
 in order to check the formula (\ref{bray_law}). 

We see two different regions that we mark with
two linear fits. The first region (left part of the figure) has a slope 
($-0.11(3)$)\footnote{
        Result of a least square fit using the points second to 
        fourth in figure 2 (seen left to right).} 
which  agrees, is a two standard deviation, with the naive 
theoretical prediction ($(\log p) A =(1.5(1) \times \log 
\frac{8}{9})=-0.18(1)$), where we have used for $A$ 
the numerator of the fit (\ref{scal}). 
The second
region decays with a behavior compatible with the equation (\ref{bray_law})
 but the slope is wrong (slope= $-0.94(4)$)\footnote{
        Using the points sixth  to 
        ninth in figure 2 (seen left to right).}. 
We think
that this decay is due to a finite--size effect (the lattice size
is 8) and hides the  decay with the ``naive'' slope ($\approx -0.18$).

Bray \cite{BRAY} shows that the real slope (in absolute value) has as
upper bound the ``naive value'' (0.18). Our numerical results go in this
direction. In particular as $T\rightarrow T_c(p)^+$ the real slope,
 in absolute value, goes to zero however the ``naive'' value will  clearly
be different from zero.

\begin{figure}[htbp]
\begin{center}
%\addvspace{1 cm}
\leavevmode
\epsfysize=250pt
\epsffile{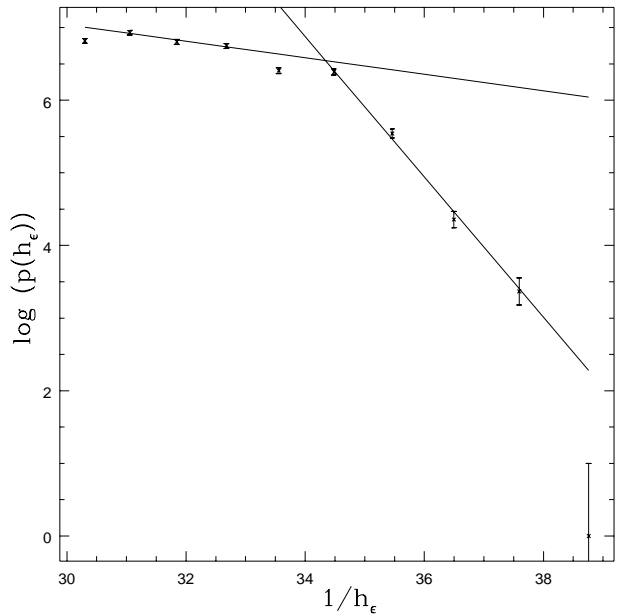}
\end{center}
\caption{Head of the logarithm of the  probability distribution 
(modulo a normalization factor) of 
the smallest zeroes for $L=8$ as a function of $1/h_\epsilon$. }
%\label{fig:chi}
\end{figure}

Hence, the numerical picture is as follows (we remark that we are in
the Griffiths phase): we have a narrow probability distribution with
its mean value having a non zero thermodynamic limit. But the minimum
value of this probability distribution follows the law of the pure
Ising model in the ferromagnetic phase so that goes to zero and
introduces a singularity in the free energy. We have seen this
behavior when simulating a large number of samples up to
$L=12$. Using a very large number of samples could be possible to
continue this result to large lattices ($L\ge 16$).

We will see in the next section how, at the critical point, the mean
value of the smallest zeroes goes to zero following a power
law.

%123456789%123456789%123456789%123456789%123456789%123456789%123456789
\section{\protect\label{S_TC}Scaling of the Yang-Lee zeroes at $T_c$}

At $T_c$ we have performed numerical simulations using the Wolff
algorithm with two degree of dilution, $p=0.889$ and $p=0.75$,
and lattice sizes $L=64, 128, 192$ and $256$. We report in Table 1
the number of samples used.

\begin{table}
\centering
\begin{tabular}{|c|c|c|} \hline
$L$ &  $p=0.889$  &  $p=0.75$ \\ \hline \hline
64  &   100       &   100     \\ \hline
128 &    40       &    40     \\ \hline
196 &    30       &    40     \\ \hline    
256 &    30       &    40     \\ \hline
\end{tabular}
\caption{
Number of samples simulated for different sizes 
and dilutions used in the numerical simulations of  sections 4 and 5.}
\end{table}

 We have used the values of the inverse critical temperatures
($\beta_c(p)$) reported in reference \cite{KIMPA}
\footnote{
        {\it i.e.} $\beta_c(p=0.889)=0.5380(3)$ and 
        $\beta_c(p=0.75)=0.772(1)$.}. 
We
will also compare the results of this reference with our
result for the $\eta$ exponent.

\begin{table}
\centering
\begin{tabular}{|c|c|c|c|c|c|} \hline
$p$  &$\gamma/\nu$&$\eta$&$\gamma/\nu$&$\Delta/\nu$&$\eta$ \\ \hline \hline
0.889&1.72(1)&0.279(14)&1.75(2) & 1.873(13)& 0.254(26)\\ \hline \hline
0.75 &1.72(3)&0.28(3)  &1.76(3) & 1.89(2)  & 0.22(4) \\ \hline
\end{tabular}

\caption{Results for the critical exponents. The first column is the
proportion of spins. The next two columns are
the critical exponents reported in reference [8]. The
fourth and fifth ones are our estimates of $\gamma/\nu$ (as control,
calculated as $\chi_{\rm max} \approx L^{\gamma/\nu}$) and
$\Delta/\nu$ (using the scaling of the zeroes) respectively. 
In the last column we report 
$\eta$ calculated using the scaling relation $\eta\equiv4-2 \Delta/\nu$.}
\end{table}

We have measured the susceptibility, 
$$
\chi=\frac{1}{V} \overline{\langle M^2 \rangle},
$$ 
where $\overline{(\cdot \cdot)}$ is the average on the disorder and 
$\langle(\cdot \cdot)\rangle$ is the thermal average. We have also
measured $\langle \cos(h M)\rangle$ in every sample to calculate the zeroes. 

We obtain $h_\epsilon$, the smallest zero for each sample, and then we
calcule the mean value, ${\overline h_\epsilon}$. 
The error is estimated using
sample to sample fluctuations. We plot the finite--size scaling in
Figure 3 and Figure 4, for $p=0.889$ and $p=0.75$ respectively, with
our best power fits drawn as a line (fifth column of Table 2).
We report the numerical values of the fit (also for the
susceptibility) in Table 2. The second and third columns of the Table
2,   are the estimates of reference \cite{KIMPA} for $\gamma/\nu$ and
$\eta$, obtained as $2-\gamma/\nu$.

\begin{figure}[htbp]
\begin{center}
%\addvspace{1 cm}
\leavevmode
\epsfysize=250pt
\epsffile{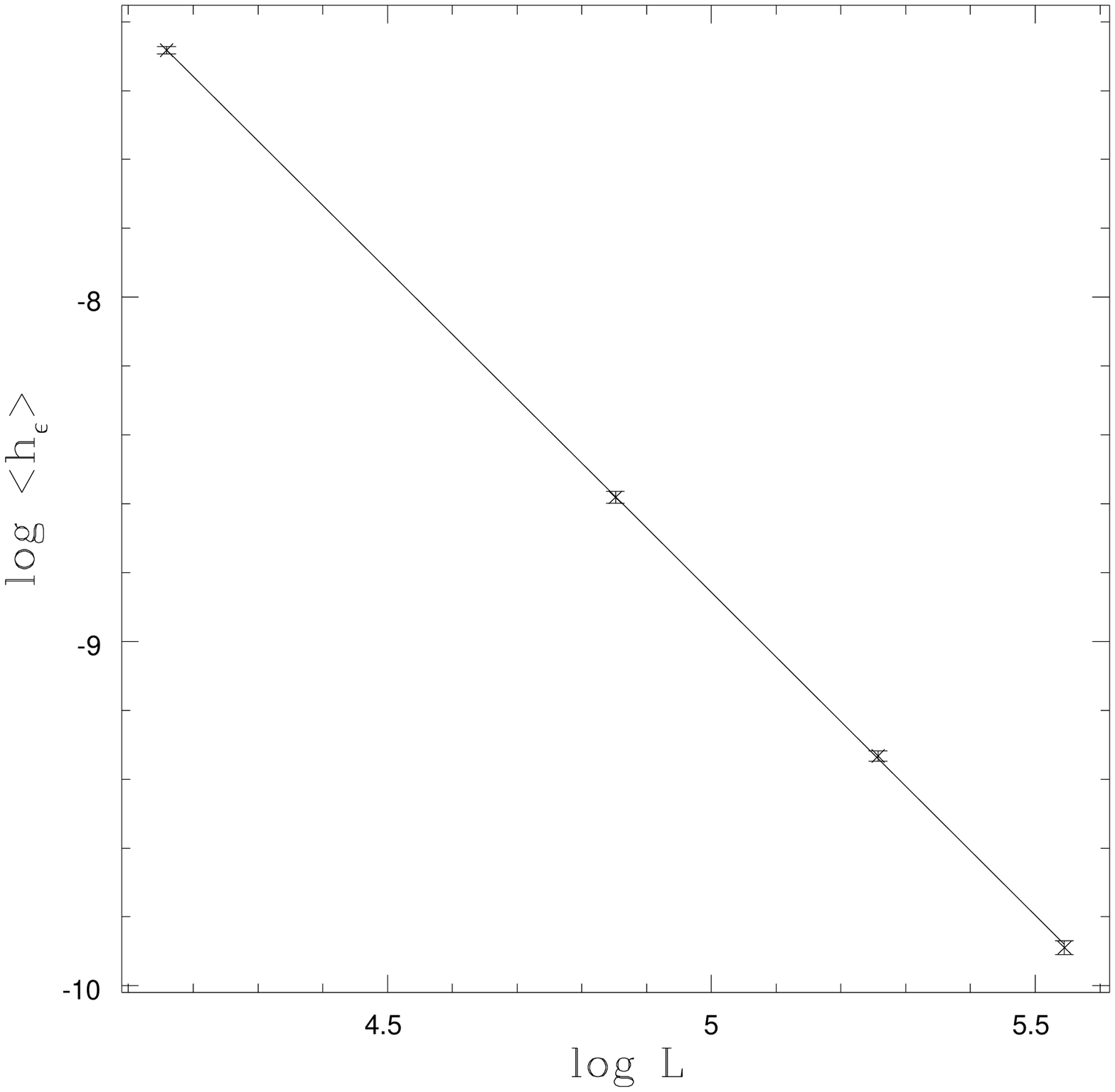}
\end{center}
\caption{Mean value of the smallest Yang-Lee zeroes,$<h_\epsilon>\equiv 
\overline h_\epsilon $, against the size, in a 
double logarithmic scale for a dilution $p=0.889$. 
The straight line is the power law fit reported in 
Table 2 and text. This fit has a  $\chi^2/{\rm DF}=0.11$. }
%\label{fig:chi}
\end{figure}

 Table 2  shows that our values of $\gamma/\nu$ are in the
errors with those of reference \cite{KIMPA} (we perform this as check)
and this also holds with our estimate of $\eta$ using scaling of
zeroes. The results are compatibles with $\eta=0.25$ on the critical line.

\begin{figure}[htbp]
\begin{center}
%\addvspace{1 cm}
\leavevmode
\epsfysize=250pt
\epsffile{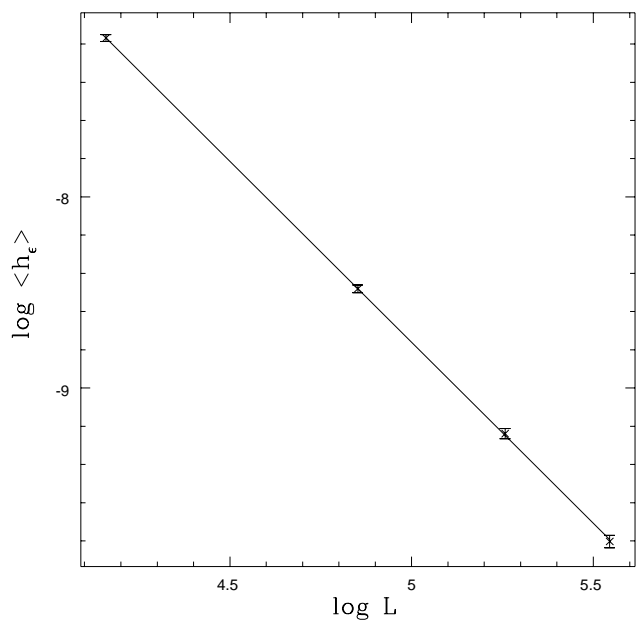}
\end{center}
\caption{Mean value of the smallest Yang-Lee zeroes against the size in a 
double logarithmic scale for a dilution $p=0.75$. 
The straight line is the power law fit reported in 
Table 2 and text. This fit has a  $\chi^2/{\rm DF}=0.28$. }
%\label{fig:chi}
\end{figure}

%123456789%123456789%123456789%123456789%123456789%123456789%123456789
\section{\protect\label{S_CON}Conclusions}

We have investigated the Griffiths phase by studying the behavior of the
probability distribution of the smallest Yang-Lee zeroes. We have
obtained a clear numerical picture of the finite--size construction of
these singularities. We have also confronted our numerical data with
previous analytical results \cite{BRAY} and the agreement is very good.

In the second part of this paper we have shown that the study of the
smallest zeroes is very useful to estimate accurately the anomalous
dimension of the system.

We have extracted one critical exponent of the
system, $\eta$, which agrees with the analytical predictions and 
with the numerical results. 
We need to calculate the second one in order to fix the
universality class of the Hamiltonian. A possible calculation, in the
line to seek complex singularities, is the study of the Fisher
zeroes \cite{SDM}. This study will point out the thermal critical
exponent $\nu$ \cite{NOW} and clarify if it depends on the proportion of
spins or not.

\section*{Acknowledgments}

We acknowledge useful discussions with F. Cesi, Vl. Dotsenko, 
A. Drory,  M. Ferrero, D. Lancaster, E. Marinari,
R. Monasson, G. Parisi.  The author is supported
by an EC HMC (ERBFMBICT950429) grant.  

\hfill

\end{document}